# Structured references from PDF articles: assessing the tools for bibliographic reference extraction and parsing

Alessia Cioffi[1] [0000-0002-9812-4065] and Silvio Peroni[2,3] [0000-0003-0530-4305]

[1] Digital Humanities and Digital Knowledge, Department of Classical Philology and Italian Studies, University of Bologna, Bologna, Italy
`alessia.cioffi@studio.unibo.it`
[2] Research Centre for Open Scholarly Metadata, Department of Classical Philology and Italian Studies, University of Bologna, Bologna, Italy
`silvio.peroni@unibo.it`
[3] Digital Humanities Advanced Research Centre (/DH.arc), Department of Classical Philology and Italian Studies, University of Bologna, Bologna, Italy

**Abstract.** Many solutions have been provided to extract bibliographic references from PDF papers. Machine learning, rule-based and regular expressions approaches were among the most used methods adopted in tools for addressing this task. This work aims to identify and evaluate all and only the tools which, given a full-text paper in PDF format, can recognise, extract and parse bibliographic references. We identified seven tools: Anystyle, Cermine, ExCite, Grobid, Pdfssa4met, Scholarcy and Science Parse. We compared and evaluated them against a corpus of 56 PDF articles published in 27 subject areas. Indeed, Anystyle obtained the best overall score, followed by Cermine. However, in some subject areas, other tools had better results for specific tasks.

## 1 Introduction

In past decades, the academic publishing world has needed to face an exponential increase in the volume of scientific literature materials [13] [29]. The necessity to handle such a vast amount of information has been one of the drivers of the digitalisation of literature materials. The conversion of academic knowledge to structured and machine-readable formats revealed positive effects also in the searchability and availability of such information, thanks to services like search engines [19]. At the same time, the structured format allowed us to valorise the citation graph connecting the scientific literature [7]. Also, in the past 50 years, bibliographic references have assumed a more prominent role in the scientific community, not only for tracking evolution in science but also for measuring impact [14].

In the past five years, the Initiative for Open Citations (I4OC, https://i4oc.org) has emphasised the importance of making citation data public. One of the main challenges to address for reaching this goal concerns extracting them from unstructured documents, like PDFs, and converting them into structured data in specific formats (e.g. JSON, XML, RDF). However, such extraction is made even more complex by the variety of (either standard or ad hoc) reference styles [15].

In the past, several tools have been proposed to address this task. Our work aims to analyse the current availability of these tools to identify which outperforms the others in extracting and parsing bibliographic references of academic papers.

The rest of the paper is structured as follows. In Section 2, we introduce the methodology adopted for identifying relevant tools and analyse their performance against a gold standard. The outcomes of the tools are shown in Section 3 and are discussed in more detail in Section 4. In Section 5, we introduce some of the essential related works in reference extraction approaches and tools. Finally, Section 6 concludes the work by sketching out some future developments.

## 2 Materials and Methods

We devised a methodology for the identification and evaluation of the reference extraction tools, which is based on four steps: (a) systematic literature review, (b) creation of a dataset, (c) creation of translation scripts, and (d) evaluation scripts. Following [31], a specific procedure was implemented and formalised in a protocol fully described in [5] – which is not reported entirely here for page constraints. Such a protocol is based on a citation-based search strategy [30] and uses seed papers for starting the search process [18]. In the first step (a), we decided to consider only papers written in English and dated after 2005. Once relevant articles were chosen in the literature, the focus moved to identify the reference extraction tools described in such documents. We decided to consider, in the analysis, only the tools that can parse full-text PDF papers, retrieve singularly tagged references, retrieve the metadata of each reference, and be either a standalone application or a programming language library, including APIs. At the end of this step, we have identified the following tools: Anystyle (https://github.com/inukshuk/anystyle-cli), CERMINE [26], EXCITE [17], GROBID [20], PDFSSA4MET (https://github.com/eliask/pdfssa4met), Scholarcy [8], and Science Parse (https://github.com/allenai/science-parse).

The next step (b) concerned preparing the data to use to test the tools identified. An initial dataset of papers in PDF format was selected to be processed by the reference extraction tools to obtain these data. This dataset included academic papers from different research fields from a corpus of selected articles used in a complementary study [24]. The dataset comprised 2,538 bibliographic references referring to almost 1,000 different journals, extracted from two articles for each one of the following 27 subject areas: Agricultural and Biological Sciences (AGR-BIO-SCI), Arts and Humanities (ART-HUM), Biochemistry, Genetics and Molecular Biology (BIO-GEN-MOL), Business, Management and Accounting (BUS-MAN-ACC), Chemical Engineering (CHE-ENG), Chemistry (CHEM), Computer Science (COM-SCI), Decision Sciences (DEC-SCI), Dentistry (DEN), Earth and Planetary Sciences (EAR-PLA-SCI), Economics, Econometrics and Finance (ECO-ECO-FIN), Energy (ENE), Engineering (ENG), Environmental Science (ENV-SCI), Health Professions (HEA-PRO), Immunology and Microbiology (IMM-MIC), Materials Science (MAT-SCI), Mathematics (MAT), Medicine (MED), Multidisciplinary (MUL), Neuroscience (NEU), Nursing (NUR), Pharmacology, Toxicology and Pharmaceutics (PHA-TOX-PHA), Physics and Astronomy (PHY-AST), Psychology (PSY), Social Sciences (SOC-SCI), Veterinary (VET). These were complemented with additional two articles having bibliographic references not introduced in a 'References' or 'Literature' section (Z-NOTES-TEST). We created a gold standard for comparing the outcomes of the reference extraction tools from these papers. We used the common metadata defining bibliographic references according to the analysis run in [24] as a baseline to understand which metadata must be identified and marked in each bibliographic reference depending on the type of the cited object.

The following step (c) consisted of translating the output of the reference extraction tools into the same format (TEI was chosen) to enable automatic comparison of such output with the gold standard. Finally, we evaluated (d) the tools using precision, recall and f-score, according to the following dimensions (based on prior studies [12] [27]):

1. *Correctly identified references*. The software's ability to distinguish each reference from the surrounding text and other references. The aim is to determine how many references are correctly identified by each parser.

2. *Correctly identified fields per reference*. The number of correctly tagged metadata, independently from content correctness. This analysis allows us to check the tools' quality of the markers' usage.

3. *Correctly identified contents per reference*. How many parts of the bibliographic reference have been correctly parsed and tagged for verifying if the text inside a correctly identified metadata is correct.

The software and all the data used for the experiment are available in [3] and [4].

## 3 Results

The overall results of the tools' assessment, introduced in Table 1, showed that Anystyle had the best performance. Nonetheless, it is possible to see a different distribution of the values between references, metadata and contents. As expected, the lowest f-score was retrieved in the correct identification of references since it was derived from the correct identification of the metadata elements and their content. Cermine showed its lowest f-score in the references dimension and its highest f-score in the metadata element identification. Overall, the dimension related to metadata contents showed that, even if the metadata element was correctly identified, the content it contained was prone to parsing errors.

The results per subject area, summarised in Fig. 1, differed slightly from the overall ones. Indeed, Anystyle showed coherent results, with all f-values above 0.5 and the highest value registered at 0.97 (BUS-MAN-ACC in Fig. 1). Another noticeable aspect is the high quality of the identification of references in the set of files which included bibliographic references in a section not labelled as "References" or "Literature" (Z-NOTES-TEST), whose p-value lay above 0.85.

**Table 1.** Precision (P), recall (R) and f-score (F1) of each dimension analysed per tool.

| Tools | References | | | Metadata | | | Content | | |
|---|---|---|---|---|---|---|---|---|---|
| | *P* | *R* | *F1* | *P* | *R* | *F1* | *P* | *R* | *F1* |
| Anystyle | **0.81** | 0.74 | **0.77** | 0.93 | **0.97** | **0.95** | 0.87 | 0.91 | **0.89** |
| Cermine | 0.75 | 0.67 | 0.71 | 0.94 | 0.94 | 0.94 | 0.86 | 0.87 | 0.86 |
| ExCite | 0.59 | 0.53 | 0.56 | 0.93 | 0.92 | 0.92 | 0.79 | 0.79 | 0.79 |
| Grobid | 0.54 | 0.55 | 0.54 | 0.86 | **0.97** | 0.91 | 0.81 | **0.92** | 0.86 |
| Pdfssa4met | 0.01 | 0.14 | 0.07 | 0.01 | 0.29 | 0.14 | 0.01 | 0.19 | 0.09 |
| Scholarcy | 0.62 | **0.78** | 0.69 | **0.96** | 0.70 | 0.81 | 0.90 | 0.65 | 0.75 |
| Science Parse | 0.43 | 0.32 | 0.37 | 1.00 | 0.55 | 0.71 | **0.94** | 0.51 | 0.66 |

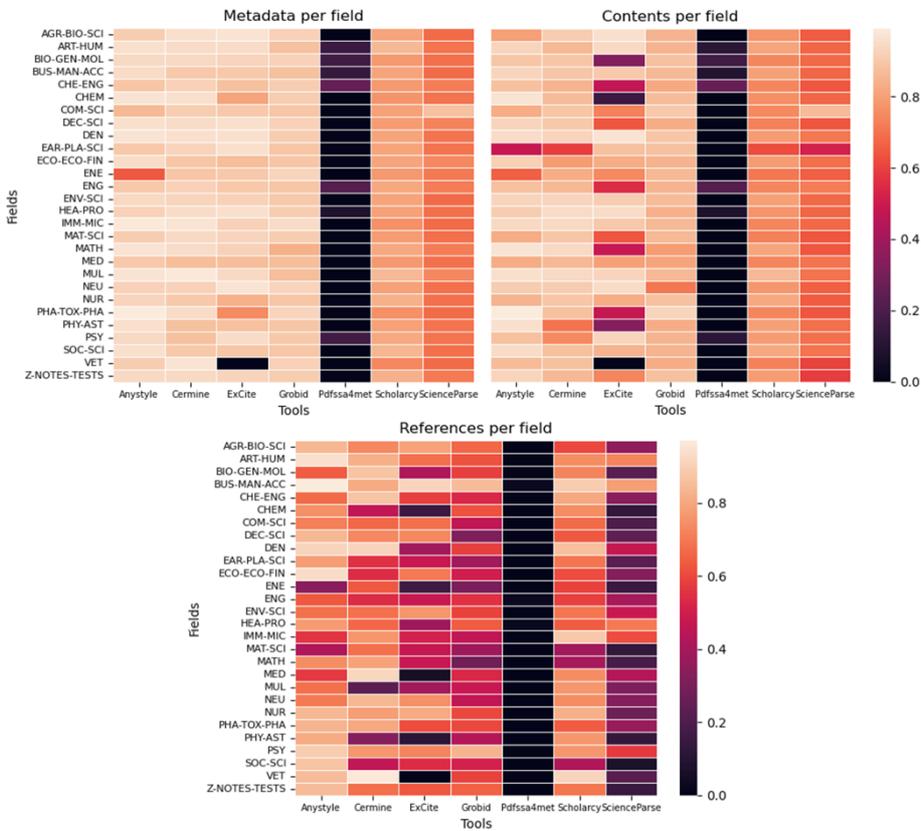

**Fig. 1** Comparison of the f-scores per subject areas (i.e. fields) in references identification.

Also, Cermine showed a high precision in reference identification, with a maximum score of 0.96 and a minimum of 0.23. The values were distributed among the fields so that, while only a few fields presented high values above 0.9, many of the fields were close to slightly lower values ranging between 0.6 and 0.8.

ExCite showed high f-scores for reference identification (e.g. 0.91 in BUS-MAN-ACC), with related high f-scores in metadata and content identification (e.g. 0.98 and 0.97, respectively, for the same subject area). However, it could not identify any bibliographic reference in the articles in VET.

The f-scores gathered using GROBID varied a lot in assessing reference identification (from 0.28 to 0.85), but they showed a smaller range in identifying contents (from 0.71 to 0.93). Pdfssa4met, instead, was the tool showing the worst performances. It was able to identify a few references (and related metadata) only in seven subject areas and showed a very low precision (from 0.01 to 0.03).

Scholarcy's f-scores highlighted the excellent performances of the tool in the main part of the subject areas, where the f-scores for the identification of references, metadata element and related content were greater than 0.58, 0.87 and 0.75, respectively. Finally, Science Parse had 0.78 as the maximum f-score in the task of reference identification (in the BUS-MAN-ACC subject area). It is worth mentioning that the precision was 1.0 in all the fields. This was not unexpected since this tool could identify only four metadata elements in each reference (i.e. author, title, source and year), thus reducing the chances of mismatching different elements.

## 4 Discussion

The comparison between tools' output and the gold standard showed a complex scenario in which a tool, Anystyle, outperformed the others. Indeed, Anystyle obtained the best score in all three dimensions of the analysis, i.e. references, metadata and contents, followed by Cermine. The remaining tools showed good performances on average, except Pdfssa4met.

It is worth mentioning that other factors that affected the reference extraction by the tools were the citation practice of particular subject areas since it affected the results mainly due to the different writing and collecting references practices. Indeed, reference identification was very effective in some subject areas, but other areas (e.g. ENE) showed low performance in all the tools. Thus, it came out that the tools' performances are affected by the practices in the subject areas and that none of the tools was good per se in all the subject areas.

This work presents three major limitations. First, the input dataset was small, even if appropriate to run initial experiments on the topic. Indeed, even if providing a vast number of research fields, each subject area included only two papers, enough to provide a preliminary insight rather than a definitive view on the topic. Second, the tools have been used off-the-shelf, without any training. For the CRF-based tools, this lack of training could have resulted in a loss in performance for some of the tools [28]. Finally, we adopted the Levenshtein distance as a unique metric to compute the similarity of the metadata content in the bibliographic references. Nonetheless, other works have identified other measures, e.g. the soft TF-IDF [6], to outperform the Levenshtein distance in measuring the similarity between two names in text retrieval tasks.

## 5 Related works

Apart from the tools identified and used in our analysis, we took notes about other theoretical approaches and workflows presented in other articles when we identified the tools to use in our study. This section presents some of the most important ones, organised in three categories.

*Single reference parsing*. This category of tools represents a set of tools which can parse a single reference and returns the metadata it is composed of in a structured format. The tools can be different depending on the approach they are based on, the input data they accept, the focus on different types of citation, e.g. academic or generic references, or the ability to extract a different number of metadata from the reference strings. Some of these tools are based on machine learning techniques, e.g. [33], while others use Hidden Markov Model [9] [32] [22], rule-based methods [25] and frame-based approaches [10] to address the same tasks.

*Parsers for reference lists.* This is a category of tools that extract and parse references from files in different formats, but not from full-text pdf files. Indeed, in most cases, they can, given a text file with a list of references (one line per reference), extract single references, parse them and return the metadata of each reference, such as Neural Parscit (https://github.com/WING-NUS/Neural-ParsCit).

*Frameworks for parsing bibliographic references in PDF full text*. In [23], the authors describe a machine-learning-based framework that outperforms the results obtained on the same input dataset by an HMM-based method. Similarly, in [26], the authors explore a composed tool based on simple HMM and rules thought to be easily modifiable by the user. Other solutions are based on rules, e.g. [1] [11] [16], ontologies [21], or deep pully convolutional networks [2].

## 6 Conclusions

This work aimed to retrieve from the available literature all the tools able to extract the bibliographic references from full-text PDF papers and evaluate them. Seven tools have been selected: Anystyle, Cermine, ExCite, Grobid, Pdfssa4met, Scholarcy and Science Parse. Three dimensions have been analysed for each: the correctly extracted metadata, the related correctly extracted contents, and the correctly extracted references.

Anystyle outperformed the others in all the three dimensions considered in the analysis. Nonetheless, the results for the analysis per subject area showed that, in some cases, Anystyle was outperformed by other tools. Thus, while Anystyle is the best tool for bibliographic reference extraction and parsing, cooperation between the tools based on the specific subtasks may be relevant to obtaining the best possible results.

In future developments, extending the current corpus of input PDF documents could be appropriate to consolidate the results obtained in this research.

**Acknowledgements.** The work of Silvio Peroni has been partially funded by the European Union's Horizon 2020 research and innovation program under grant agreement No 101017452 (OpenAIRE-Nexus).